\documentclass[aps,prb,twocolumn,showpacs,floatfix]{revtex4-1}
\usepackage{graphicx}
\usepackage{color}
\usepackage{amsmath}
\usepackage{amssymb}
\usepackage{amsfonts}
\definecolor{scarred}{rgb}{0.75,0.0,0.0}
\usepackage{hyperref}
\hypersetup{
colorlinks=true,final=true,
        linkcolor=blue,
        citecolor=blue,
        filecolor=blue,
        urlcolor=blue,
}

\begin{document}
\title{Photo-excited states in correlated band insulators}
\author{Nagamalleswararao Dasari$^{1,2}$}\email{nagamalleswararao.d@gmail.com}
\author{Martin Eckstein$^{1}$}
%\affiliation{$^{1}$ Max-Planck institute for the structure and dynamics of matter, 22761 Hamburg, Germany.}
\affiliation{$^{1}$University of Erlangen-Nuremberg, 91058 Erlangen, Germany}
\affiliation{$^{2}$Max Planck Institute for the Structure and Dynamics of Matter, 22761 Hamburg, Germany.}
%%%%%%%%%%%%%%%%%%%%%%%%%%%%%%%%%%%%%%%%%%%%%%%%%%%%%%%%%%%%%%%%%%%%%%%%%%%%%%%%%%%%
%%%%%%%%%%%%%%%%%%%%%%%%%%%%%%%%%%%%

\begin{abstract}

We study the photo-excitation dynamics of correlated band insulators, 
using non-equilibrium dynamical mean-field theory for the ionic 
Hubbard model. We find two distinct behaviors, depending on the 
ratio of the on-site interaction $U$ and the bare band gap 
$\Delta$. For small interactions, the relaxation is characterized 
by intra-band carrier scattering in relatively rigid bands, 
leading to a non-thermal intermediate state with separate 
thermal distributions of electrons and holes. This behavior can be viewed as typical 
for a band insulator with weak interactions. For larger 
interaction, on the other hand, we observe a strong modification 
of the electronic spectrum and a filling-in of the gap after 
photo-excitation, along with a rapid thermalization of the 
system. The two behaviors therefore provide a dynamical 
distinction of a correlated band insulator and a band-insulator, which can differ 
even when the spectra of the two systems are similar in 
equilibrium. The crossover happens when the interaction $U$ is 
comparable to $\Delta$.

\end{abstract}
%\pacs{71.10.Fd}

\maketitle

\section{Introduction}

The possibility to switch quantum states through ultrafast 
non-thermal pathways has made non-equilibrium studies of quantum 
many-body systems attractive to the condensed matter 
community.\cite{doi:10.1080} Using laser photo-excitation 
or external electric fields one can induce transient states on short time-scales, 
and in some cases even long-lived hidden quantum states, with electronic 
properties which are  entirely different from 
equilibrium.\cite{Fausti189,Stojchevska177,Vaskivskyie1500168,Nasu}

A rich variety of novel phases in non-equilibrium can be expected for 
strongly correlated systems. Electronic correlations are predominantly 
studied in systems which are metallic in the non-interacting limit.
In these systems, correlations result in phenomena like the Mott 
transition, high-temperature superconductivity, and non-Fermi liquid 
behavior. In band insulators, in contrast, one may naively expect that 
the existence of the band gap and the absence of low energy quasi-particles 
reduces the importance of electronic correlations. However, 
this turns out to be not true. The role of electronic correlations in 
band insulators has been studied in various models, including the ionic 
Hubbard model,\cite{0953-8984-15-34-319,PhysRevB.75.245122,
PhysRevB.70.155115,PhysRevLett.92.246405,PhysRevLett.97.046403,
PhysRevLett.98.016402,PhysRevB.78.075121,PhysRevLett.98.046403,
PhysRevB.79.121103,PhysRevB.89.165117,hoang2010metal,PhysRevLett.112.106406} 
a two orbital Hubbard model with crystal-field splitting,
\cite{PhysRevLett.99.126405} a two sub-lattice model with 
inter-orbital hybridization\cite{PhysRevB.80.155116, PhysRevB.87.125141} 
and a bilayer model with two identical Hubbard planes.
\cite{PhysRevB.59.6846,PhysRevB.73.245118,PhysRevB.75.193103,
PhysRevB.76.165110,0295-5075-85-3-37006} In general, at weak-coupling 
the competition between the local Coulomb interaction and the 
non-interacting band gap results in a correlated band insulator 
with a renormalized gap. Further increasing the interaction 
strength can even close the gap, and lead to an interaction-driven 
metal or bond-ordered state in the strong coupling limit.
\cite{PhysRevLett.97.046403,PhysRevLett.98.016402,PhysRevB.80.155116} 
The renormalized band gaps and the difference in charge and spin gaps 
distinguishes correlated band insulators from non-interacting band 
insulators.\cite{0953-8984-15-34-319,PhysRevB.75.245122,PhysRevB.80.155116} 
Correlated band insulators provide a paradigm example for systems with 
competing ground state interactions, and hence are good candidates to 
explore hidden quantum states through non-thermal pathways.

In a band insulator, photo-excitation will promptly lead to a 
partial occupation of the conduction band. At weak coupling, 
one may expect the dynamics of such systems to be similar to semiconductors: 
Interactions provide a mechanism for electron-electron scattering, while the 
bands which are almost rigid apart from some photo-induced screening 
of the gap which is largely included in a Hartree shift of the bands. 
This leads to a state in which electrons and holes are separately 
thermalized in the conduction and valence band. On longer timescales, 
the  system would establish a common temperature and chemical potential, 
and the energy is passed to the lattice. Such processes in semiconductors 
are well described using quantum Boltzmann equations.\cite{Haug} In this 
paper, we explore how this picture is modified at larger interaction, 
and whether there is a distinct dynamical behavior of band insulators 
and correlated band insulators. How is the gap renormalized or even filled 
due to local correlations after photo-excitations? What is the role of 
these correlations on the thermalization process: Does the filling in of 
the gap lead to a speed-up of thermalization, and is there room for 
non-thermal metallic states? 

The theoretical study of such systems out of equilibrium is 
challenging. Kinetic equations cannot describe a change of the 
electronic structure due to local correlations. 
In the past few years, non-equilibrium dynamical mean-field 
theory (DMFT) has been widely applied to study the photo-excitation and 
quench dynamics of correlated electronic systems.\cite{RevModPhys.86.779} 
In this work, we study the dynamics of correlated band 
insulators by means of non-equilibrium DMFT, using iterative perturbation 
theory (IPT) as an impurity solver. We find a crossover between two 
different behaviors when the local Coulomb interaction becomes 
comparable to the non-interacting band gap. The typical behavior of a band 
insulator with two separate sub-systems of thermalized holes 
and electrons  persists only for the smaller interaction. 

The paper is  organized as follows. In section \ref{sec2}, we 
present the model and method used to study 
the photo-excitation of correlated band insulators, and 
briefly discuss the equilibrium properties of the model. In 
section \ref{sec4} we present the numerical results, and 
Sec.~\ref{sec:discussion} contains an analysis and discussion.
             
\section{Model and Method}
\label{sec2}

\subsection{Ionic Hubbard model and DMFT}
\label{sec2a}

Correlated band insulators are well described by the ionic 
Hubbard model (IHM) on a bipartite lattice with two sub-lattices 
$A$ and $B$. The Hamiltonian is given by
\begin{align}
{\mathcal{H}}= & -J(t) 
\sum_{\langle i,j\rangle}[c^{\dagger}_i
%\sum_{i\in A, j\in B}[c^{\dagger}_i
 c^{\phantom \dagger}_j + H.c] + \Delta \sum_{i\in A} n_i - 
\Delta \sum_{i\in B} n_i \nonumber \\& + 
U \sum_i n_{i\uparrow} n_{i\downarrow} - \mu \sum_i n_i, 
\label{eq:model}
\end{align}
where $J$ is the nearest neighbour hopping amplitude, and $U$ is the 
on-site Coulomb interaction. The staggered ionic potential $\Delta$ is 
the origin of the non-interacting band gap. We choose the chemical 
potential $\mu$ such that total filling of the system is fixed at 
half-filling ($n_A+n_B$=1). At half-filling, the above Hamiltonian 
describes a band insulator and a Mott insulator in the limits 
($U$ = 0, $\Delta \neq $0) and ($U$ = $\infty$, $\Delta$ = 0), 
respectively. Nontrivial states arise when both $\Delta$ and $U$ are 
nonzero and finite.

To study the photo-excitation dynamics of correlated band 
insulators we employ the non-equilibrium DMFT formalism,
\cite{RevModPhys.86.779} which is exact in the limit of infinite-coordination 
number ($Z$) with hopping $J(t) = J^*/\sqrt{Z}$. We choose a 
non-interacting dispersion of bands on a bipartite lattice, 
with a semi-elliptical density of states $\rho(\epsilon) = 
\frac{1}{\pi J^*} \sqrt{4{J^*}^2 -\epsilon^2}$. We set the half bandwidth $2J^*=1$ 
as energy unit, and the unit of time is $\frac{\hbar}{2J^*}$. The 
DMFT formalism makes a local approximation for the self-energy, 
i.e., $\Sigma_{ij}(t,t') \approx \Sigma_{ii}(t,t')$ which 
allows to map the lattice problem Eq.~\eqref{eq:model} to an 
effective quantum impurity problem with a self-consistently 
determined bath $\Gamma(t,t')$. The impurity action for 
sub-lattice $\alpha\in\{A,B\}$ is given by
\begin{equation}
{S_{\alpha}} = \int_C dt\ H^{\alpha}_{loc}(t) + \sum_{\sigma} 
\int_C dtdt'\ c^{\dagger}_{\sigma\alpha}(t) \Gamma_{\alpha}(t,t') c_{\sigma\alpha}(t).
\label{eq:action}
\end{equation}
Here $C$ is the Keldysh time contour (for an 
introduction into the Keldysh formalism, 
see, e.g., Ref.~\onlinecite{RevModPhys.86.779}), and
\begin{equation}
H^{\alpha}_{loc} = U n^{\alpha}_{\uparrow} n^{\alpha}_{\downarrow} 
+ [\Delta_\alpha-\mu] (n^{\alpha}_{\uparrow} +n^{\alpha}_{\downarrow} )
\end{equation}
is the local Hamiltonian on the impurity, with $\Delta_A=\Delta$ 
and $\Delta_B=-\Delta$. The semi-elliptical density of states 
yields a closed form expression for the self-consistency relation, which is given by 
$\Gamma_{\alpha}(t,t') = J^*(t)G_{\overline{\alpha}} (t,t') J^*(t')$, 
where $G_{\alpha}(t,t')= -i{{\langle {\rm{T}}_C c_{\alpha}(t) 
c^{\dagger}_{\alpha}(t')\rangle}_S}_{\alpha}$ is the local Green's function 
defined on Keldysh contour ($\overline{\alpha}=B$ for $\alpha=A$ and 
vice versa). For a given sub-lattice hybridization $\Gamma_{\alpha}$, the 
local Green's functions $G_{\alpha}$ can be determined from the Dyson equation,
\begin{equation}
[i\partial_t+\mu+\Delta_{\alpha}-\Sigma_\alpha-\Gamma_{\alpha}]G_{\alpha} = 1, 
\end{equation}
where $\Sigma_{\alpha}$ is local self-energy which is calculated from the 
effective impurity problem. We use the real-time iterative perturbation 
theory (IPT) as an impurity solver to calculate the local self-energy. The 
diagrammatic expression for the IPT ansatz is given by
\begin{align}
\Sigma_{\alpha}(t,t')=
& \sqrt{A^{\alpha}(t)A^{\alpha}(t')} U(t)U(t')  \,\,\times\nonumber\\
&\times \,\,\,
{\mathcal{G}^{H}_{\rm{\alpha}}}(t,t')
{\mathcal{G}^{H}_{\rm{\alpha}}}(t,t'){\mathcal{G}^{H}_{\rm{\alpha}}}(t',t),
\label{ipt}
\end{align}
where ${\mathcal{G}^{H}_{\rm{\alpha}}}$ is the Hartree-corrected 
bath propagator, which is given by the Dyson equation
\begin{equation}
[i\partial_t+\mu-\Delta_\alpha-\Sigma^{\alpha}_{H}-
\Gamma_{\alpha}]{\mathcal{G}^{H}_{\rm{\alpha}}} = 1, 
\end{equation}
with the Hartree self-energy $\Sigma^{\alpha}_{H}(t,t') = U n^{\alpha}(t) 
\delta_c(t,t')$. Following Ref.~\onlinecite{PhysRevLett.97.046403}, 
the factors $A^{\alpha}$ in Eq.~\eqref{ipt} are chosen such that in 
equilibrium the ansatz is exact in the weak-coupling 
limit ($U/t\ll$ 1) and it has the exact short time behavior 
for all values of $U/t$, which imposes various exact sum rules. 
The corresponding expression for $A^{\alpha}(t)$ is 
$n_{\alpha}(t)(1-n_{\alpha}(t))/n^0_{\alpha}(t)(1-n^0_{\alpha}(t))$ 
where $n^0_{\alpha}(t)$ is the impurity occupancy, 
which is calculated from ${\mathcal{G}^{H}_{\rm{\alpha}}}$. 
Writing the $A$-factor in the symmetrized form 
$\sqrt{A^{\alpha}(t)A^{\alpha}(t')}$ guaranties that the 
self-energy Eq.~\eqref{ipt} is hermitian, while the expression 
reduces to the conventional form $\sqrt{A^{\alpha}(t)A^{\alpha}(t')}=A_\alpha$ 
at equilibrium, when $A_\alpha$ is time independent.

Non-equilibrium DMFT measures the local Green's function 
for each sub-lattice. The sub-lattice occupancy is obtained 
from the equal-time lesser component of the Green's function, 
$m(t) = n_A(t)-n_B(t)=-iG^<_{A}(t,t)+iG^<_{B}(t,t)$. To find the 
effective temperature of the photo-excited system we need to 
calculate the total energy of the system, which is the sum of 
kinetic energy, interaction energy and lattice potential energy. 
The kinetic energy of photo-excited system is given by the equal-time 
contour convolution of the hybridization function with the local 
Green's function, $E_{kin}$ = $-i\sum_{\alpha}[\Gamma_{\alpha}*G_{\alpha}]^<(t,t)$, 
and the interaction energy is the equal-time contour convolution 
of the local self-energy with the local Green's function, 
$E_{int}$ = $i\sum_{\alpha}[\Sigma_{\alpha}*G_{\alpha}]^<(t,t)$.\cite{RevModPhys.86.779} 
The lattice potential energy is $E_{\Delta}(t) = \Delta (n_A(t)-n_B(t))$. The 
double occupancy $d_{\alpha}(t)$ = 
$\langle n^{\alpha}_{\uparrow}n^{\alpha}_{\downarrow}\rangle$ is 
obtained from the interaction energy. 

One of the issues with IPT is the non-conserving nature of the 
self-energy in the sense of Baym-Kadanoff. A violation of the energy 
conservation can be identified when there is a drift of the 
total energy with time even though the Hamiltonian is time 
independent. In previous non-equilibrium DMFT calculations for 
interaction quenches in a single-band Hubbard model it was found 
that IPT conserves the total energy very accurately up to all 
accessible simulation times in the weak-to-intermediate coupling 
regime, while with increasing $U$ there is a relatively sharp crossover 
after which only the short time dynamics is correctly captured.
\cite{PhysRevB.81.115131} In this work we focus on relatively weak 
coupling, and we have confirmed than an unphysical short-time increase 
of the energy as observed in Ref.~\onlinecite{PhysRevB.81.115131} does not occur.

\subsection{Equilibrium DMFT study of the IHM}
\label{sec2a}

A first understanding of the equilibrium states of the IHM can be obtained from the atomic 
limit $J/U$=0, at total filling equal to 1. 
In this trivially soluble limit, the ground state for $U<2\Delta$ 
has two electrons on sublattice $B$ and zero on sublattice $A$, resulting a band insulator with a band gap 
$\Delta-U/2$. In the opposite limit ($U>2\Delta$) each sublattice is occupied by one 
electron, and we get a Mott insulator with a gap $U$. In the atomic limit, the system 
is gapless at $U$=2$\Delta$. An interesting question therefore is whether 
local correlations broaden this metallic point into a metallic phase when $J$ is non-zero.

Equilibrium studies of the IHM at zero temperature using DMFT with an IPT impurity 
solver, find that electronic correlations strongly renormalize the non-interacting 
band-gap for small values of $U$. The 
gap in the spectral function vanishes  at a critical interaction $U_c$, and the 
system is metallic for a finite range of interactions.\cite{PhysRevLett.97.046403,
PhysRevB.78.075121} As we further increase $U>U_c$, the system 
becomes Mott insulating, i.e., the gap in the spectral function opens again and increases 
with $U$. The crossover from a correlated band insulator to a metal has been observed experimentally in 
photoemission spectra of SrRu$_{1-x}$Ti$_x$O$_3$.\cite{PhysRevB.76.165128}.

IPT studies of the IHM find the metallic phase only up to intermediate values 
of $\Delta$, while for large values of the ionic potential there is a 
direct transition from a band insulator to a Mott insulator.\cite{PhysRevLett.97.046403,
PhysRevB.78.075121} A recent study of the IHM at finite temperature using a continuous-time quantum 
Monte Carlo (CTQMC) impurity solver has obtained a phase diagram similar to 
IPT for intermediate values of $\Delta$, but in contrast to IPT the 
intermediate metallic phase persists to even larger values of 
$\Delta$.\cite{PhysRevB.89.165117} An open issue is the nature intermediate metallic 
phase.\cite{PhysRevB.78.075121}
2D-cluster extensions of DMFT find a bond-ordered phase as an intermediate 
metallic phase,\cite{PhysRevLett.98.016402} while 
single-site DMFT using CTQMC suggests that the intermediate metallic phase is a 
Fermi-liquid.\cite{PhysRevB.89.165117}

\subsection{Photo-doping}
\label{sec3}

Photo-doping changes the electron-hole concentration by irradiating light on the sample. 
Experimentally, laser photo-excitation not only changes the carrier concentration of the bands 
at the Fermi-level, but it can also excite electrons from bands far below the 
Fermi-level into the conduction band, and electrons from the valence band to 
other bands far above the Fermi-level. In our work, we employ this setup to 
induce photo-excited state. To mimic such an experimental situation 
theoretically, one can couple two additional wide-band particle 
reservoirs to the system. One reservoir is entirely filled and 
will therefore inject particles into the empty states of the system, 
while the other one is empty and can take out electrons from occupied states. 
The additional fermionic baths are non-interacting and can be integrated out 
exactly. The resulting DMFT action is the same as Eq.~\eqref{eq:action}, but 
the hybridization function  is modified to 
\begin{equation}
{\Gamma'}_{\alpha}(t,t') = \Gamma_{\alpha}(t,t') + \Gamma^{filled}(t,t') + \Gamma^{empty}(t,t').
\end{equation}
Here the first term corresponds to the self-consistent 
hybridization of the ionic Hubbard model, and the last two terms 
correspond to the hybridization of the filled and empty reservoirs. The latter are given by
\begin{equation}
\Gamma^{filled/empty}(t,t') = h(t)G^{filled/empty}(t,t')h(t'),
\end{equation} 
where $G^{filled/empty}$ is the Green's function of the 
reservoir, and $h(t)$ is the time-profile of the coupling 
between the IHM and the reservoirs, for which we take the form
\begin{equation}
h(t) = h_0\sin(\pi t/t_{pulse}),
\end{equation}
for $t \le t_{pulse}$ and $h(t)=0$ for $t > t_{pulse}$. 
Throughout this work we fix $t_{pulse}$=3.0, and $h_0$ is choosen 
to control the excitation density. The density of states 
corresponding to $G^{filled/empty}(t,t')$ is a Lorentzian of width $8J^*$. 

\section{Results}
\label{sec4}

We start the analysis of relaxation after photo-excitation 
with an investigation of the time-dependent double occupancy 
and the sublattice occupation (Sec.~\ref{sec33a} and \ref{sec33b}), and 
then proceed to analyze spectral functions and the occupation 
function (Sec.~\ref{sec33c}). To know whether the photo-excited 
system thermalizes after the pulse, we need to find the effective 
temperature $T_{\text{eff}}$ of the time-evolved state.
\cite{PhysRevB.84.035122} For this purpose, we determine a thermal 
equilibrium state which has the same energy as the time-evolved state. 
The corresponding temperature of the equilibrium state then defines $T_\text{eff}$. 
(In practice, we choose the excitation density in order to fix a given effective temperature, 
$T_\text{eff}=1$ or $T_\text{eff}=0.3$.) With this we can compare 
time-dependent expectation values of various observables $O(t)$ to 
corresponding expectation values $O_{th} \equiv O(T_{\text{eff}}) = 
Tr[\exp^{-H/T_{\text{eff}}} O]/Z$. Throughout the paper we fix a low 
initial temperature $T$ by $\beta=1/T = 32$.    

\begin{figure}[t]
\centering
\includegraphics[angle=0,width=1.0\columnwidth]{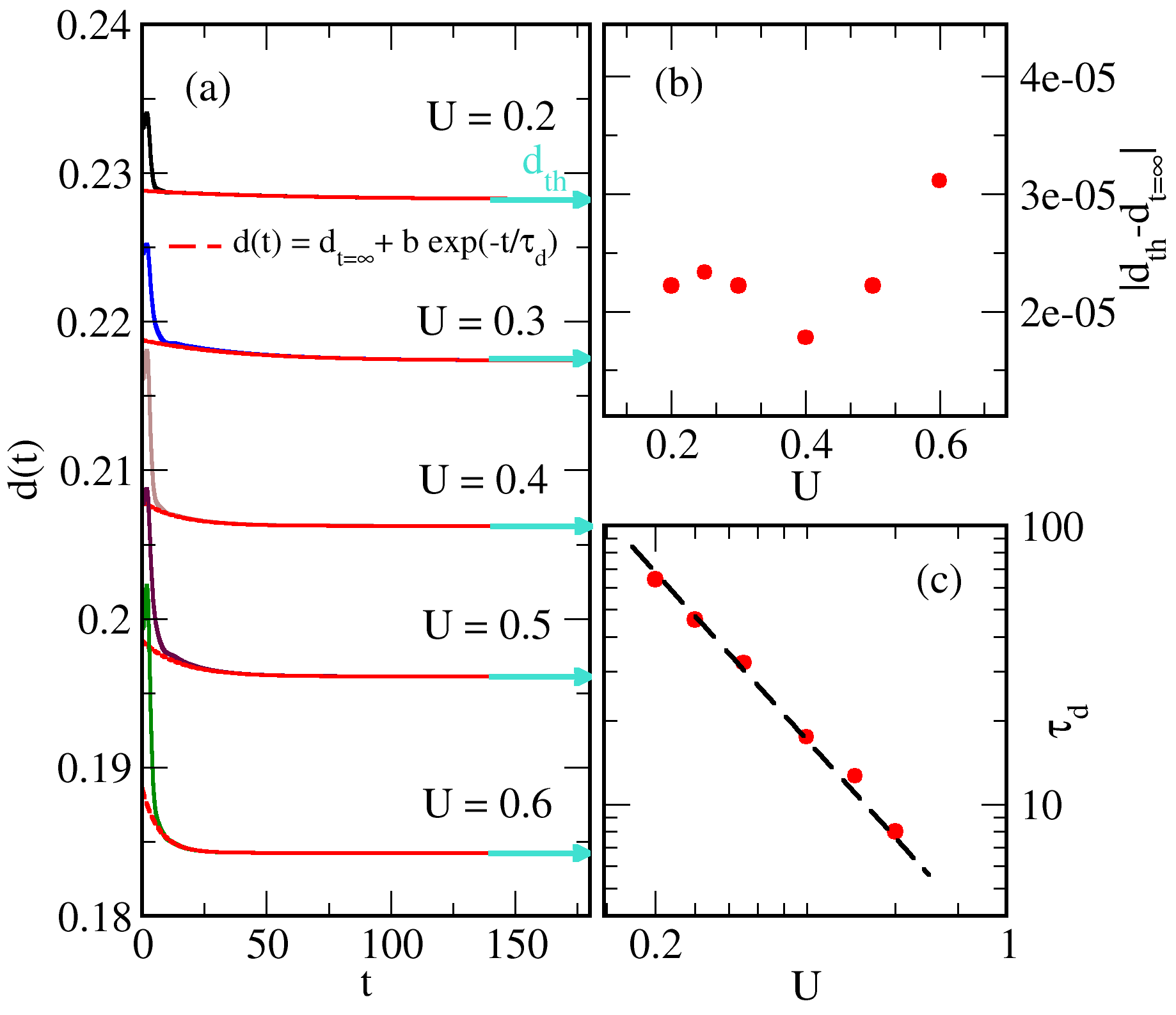}
\caption{(a) Double occupancy of the photo-excited system for different 
values of $U$ values. The excitation density is choosen such 
that $T_{\rm{eff}}$ = 1. Solid curves show the numerical data, 
and red dashed lines show exponential fits. Small arrows indicates 
the double occupancy in thermal equilibrium. (b) Difference of the 
double occupancy in equilibrium at temperature $T=T_{\rm{eff}}$ and 
in the photo-excited system. (c) Relaxation time of the double 
occupancy for different values of $U$. The dashed curve is a 
fit to the relaxation times with the relation $\frac{a}{U^2}$.}
\label{fig:fig1}
\end{figure}

\subsection{Thermalization of the metal ($\Delta= 0$)}
\label{sec33a}

When the ionic potential is zero, the IHM reduces to a simple one-band 
Hubbard model with intra-orbital interaction $U$, which is metallic at 
small values of $U$. In order to later contrast the results to the 
correlated band insulator, we first investigate the thermalization of this 
metallic phase within our formalism. In Fig.~\ref{fig:fig1}(a), we plot 
the time-dependent double occupancy $d(t) = \langle n_\uparrow n_\downarrow \rangle$ 
of the photo-excited system for different values of $U$. We fit the time-dependent 
double occupancy to a single exponential function 
$d(t)=d_{t=\infty}+b\exp(-t/\tau_d)$ and extract the extrapolated 
value $d_{t=\infty}$ of $d(t)$ in the long time limit and the 
corresponding relaxation times $\tau_d$. The thermal values $d_{th}$ of 
the double occupancy, obtained from equilibrium simulations at the 
corresponding effective temperature are shown by arrows in 
Fig.~\ref{fig:fig1}(a). The difference $|d_{th}-d_{t=\infty}|$ is 
plotted in Fig.~\ref{fig:fig1}(b). One can see that the time-dependent 
double occupancy approaches $d_{th}$ in the long-time limit for all 
values of $U$ within our numerical accuracy, which implies thermalization of 
the weakly-correlated metal. The relaxation times $\tau_d$ are plotted in 
Fig.~\ref{fig:fig1}(c). The thermalization times can be fit well with a 
power law $\tau_d \sim 1/U^2$. This behavior is consistent with a quasi-particle 
picture: In the different calculations we have chosen the excitation such that 
the effective temperature is fixed to $T_{\text{eff}}=1$, so that the phase space 
for scattering is fixed, and the scattering rate is given by the scattering 
matrix element, which is proportional to $U^2$. 
\begin{figure*}[t]
\centering
\includegraphics[angle=0,width=\textwidth]{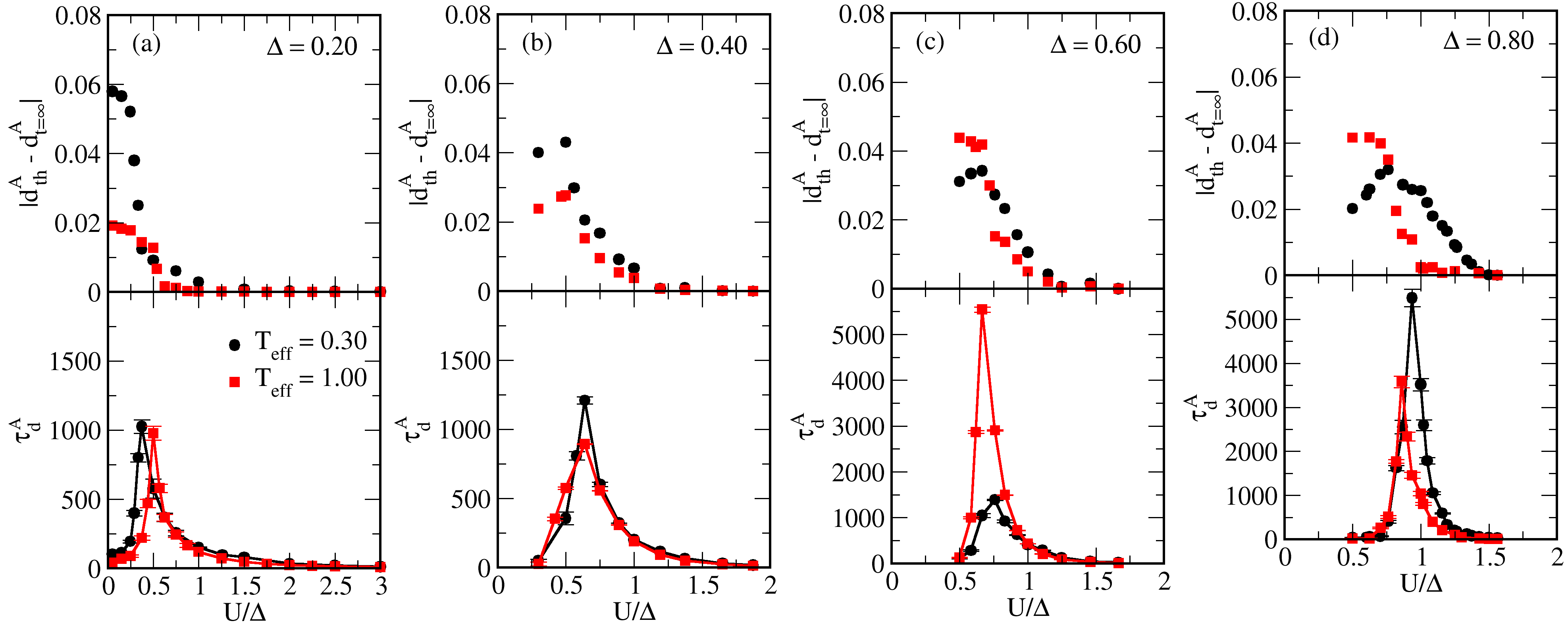}
\caption{Upper panel: Difference of the double occupancy 
in thermal equilibrium at temperature $T=T_{\rm{eff}}$ and in the time-evolved 
state, for different values of the ionic potential (a) $\Delta$ = 0.2 (b) 
$\Delta$ = 0.4 (c) $\Delta$ = 0.6 (d) $\Delta$ = 0.8. The excitation 
density is choosen such that $T_{\rm{eff}}$ = 0.3 (black circles) or 
$T_{\rm{eff}}$ = 1.0 (red squares). 
Lower panel: Relaxation times of the double-occupancy.}
\label{fig:fig2}
\end{figure*}

\subsection{Correlated band insulator: relaxation of the double occupancy and the sub-lattice occupation}
\label{sec33b}

To study the relaxation dynamics of the correlated band insulator, 
we analyze the relaxation dynamics of the time-dependent 
double occupancy $d^A(t)$ (measured at the A sub-lattice 
without loss of generality). Analogous to the previous paragraph, 
we extract the value of the double occupancy at infinite time 
($d^A_{t=\infty}$) and the relaxation time ($\tau^A_d$) from an 
exponential fit $d^A(t) = d^A_{t=\infty} + b\exp(-t/\tau^A_d)$. In the 
upper panel of Fig.~\ref{fig:fig2}, we plot the difference of $d^A_{th}$ to 
the time evolved value $d^A_{t=\infty}$ for different values of the ionic 
potential. In contrast to the metal, $|d^A_{th}-d^A_{t=\infty}|$ approaches 
zero only when $U/\Delta \gtrsim$ 1. This implies that for smaller 
interactions, the observed relaxation on the timescale of our 
simulation is towards a non-thermal state, while the behavior at 
large interactions is consistent with thermalization. The crossover 
from non-thermal to thermal behavior occurs roughly when $U$ is of the 
order of $\Delta$, but it also depends on the excitation density 
(i.e., the effective temperature of the final state). In the lower 
panel of Fig.~\ref{fig:fig2}, we plot the relaxation times $\tau^A_d$ 
obtained from the exponential fits. In the regime where the system 
thermalizes ($U/\Delta \gtrsim1$), the relaxation times decrease 
with $U$, like in the metal. In the opposite limit, the relaxation 
depends on $U$ in a non-monotonous way, and reaches a maximum value around U/$\Delta \approx$ 0.6. 

\begin{figure*}[t]
\centering
\includegraphics[angle=0,width=\textwidth]{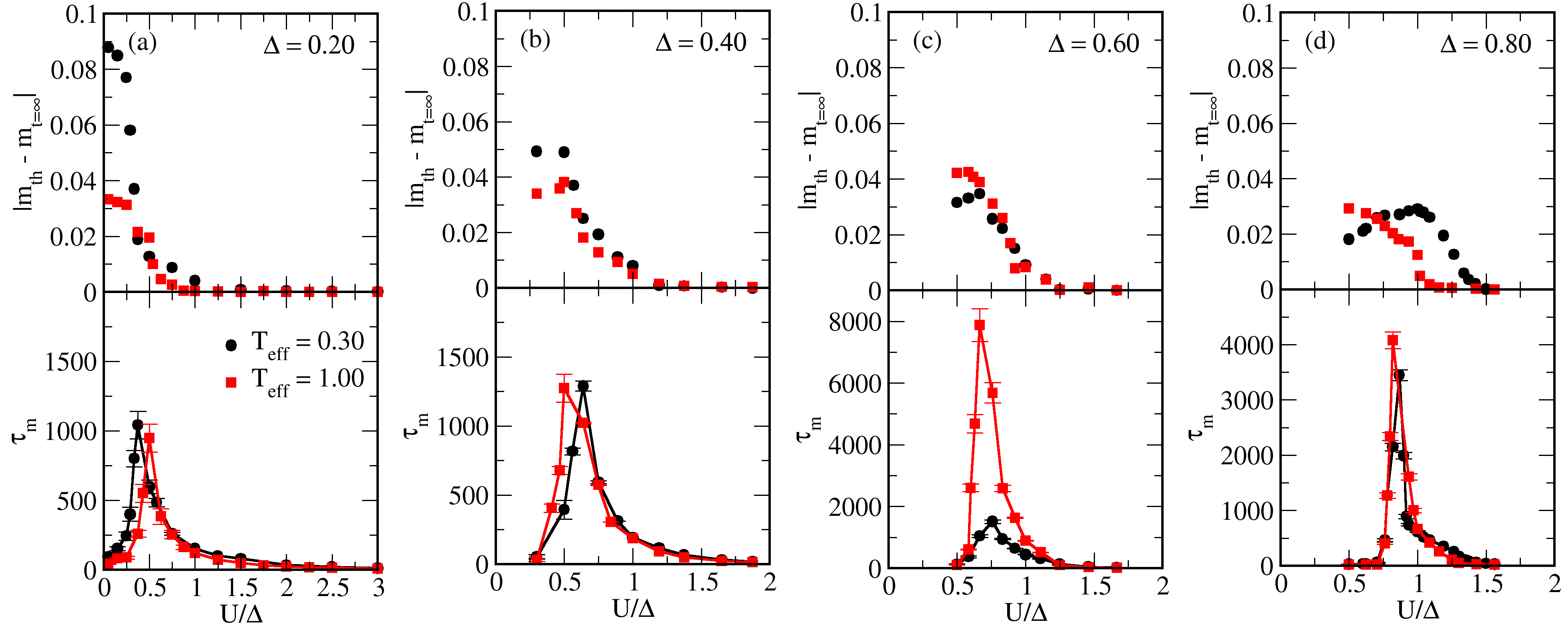}
\caption{Upper panel: Difference of the sub-lattice occupancy in 
thermal equilibrium ($m_{th}$) at temperature $T=T_{\rm{eff}}$ and 
in the time-evolved state ($m_{t=\infty}$), for different values 
of the ionic potential (a) $\Delta$ = 0.2 (b) $\Delta$ = 0.4 (c) 
$\Delta$ = 0.6 (d) $\Delta$ = 0.8. The excitation density is choosen 
such that $T_{\rm{eff}}$ = 0.3 (black circles) or $T_{\rm{eff}}$ = 
1.0 (red squares). Lower panel: Relaxation times of the sub-lattice occupancy.}
\label{fig:fig3}
\end{figure*}

The sub-lattice occupancy $m(t)=n_A(t)-n_B(t)$, exhibits a similar 
behavior as observed for the double occupancy. For this purpose, we 
again analyze $m(t)$ by means of an exponential fit, 
$m(t)=m_{t=\infty}+b\exp(-t/\tau_m)$. The upper panel of 
Fig.~\ref{fig:fig3} shows the difference of 
$m_{th}$ and $m_{t=\infty}$ for the same parameters as 
in Fig.~\ref{fig:fig2}. Similar to the double occupancy, the 
sub-lattice occupation $m(t)$ displays a crossover from non-thermal 
to thermal behavior when $U$ is comparable to the ionic potential, 
with the same trend in the relaxation times $\tau_m$  (lower panel of Fig.~\ref{fig:fig3}).

\begin{figure}[t]
\centering
\includegraphics[angle=0,width=1.0\columnwidth]{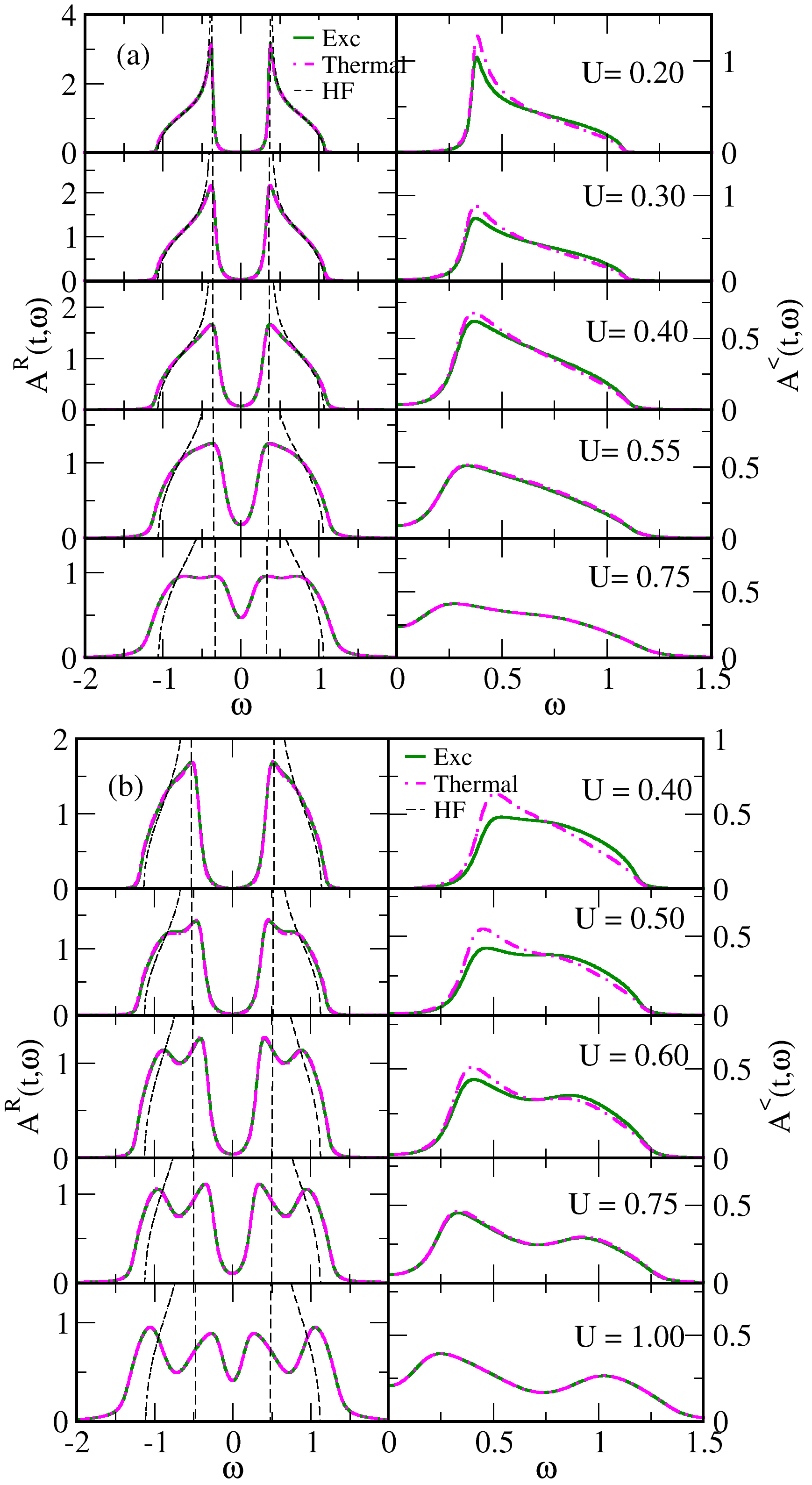}
\caption{(a) Left panel: Spectral function ${A^R(t=180,{\omega})}$ for 
different values of $U$ and $\Delta$ = 0.4. (a) Right panel: The corresponding 
occupied density of states ${A^<(t=180,{\omega})}$ above the Fermi-level. 
The black dashed lines indicate the Hartree-Fock spectral function 
(see text) (b) Same as (a) but for $\Delta$ = 0.6. 
The excitation density is choosen such that $T_{\rm{eff}}$=1.}
\label{fig:fig4}
\end{figure}

\subsection{Relaxation of the spectral function}
\label{sec33c}

The observed crossover from non-thermal to thermal behavior can be 
further analyzed by looking at the spectral functions and the 
occupation function. The analysis of the spectral function 
furthermore allows to access the possible transient metallicity of 
the photo-doped state. To identify any such metallic states we calculate 
the single particle spectral function $A^R(\omega,t)$ and the occupied 
density of states $A^<(\omega,t)$ from a Fourier transform,
\begin{equation}
\label{ftspec}
A^{\gamma}(t,\omega) = \mp \frac{1}{\pi} {\rm{Im}} \int^{t}_0 ds e^{i\omega s} G^{\gamma}(t,t-s).
\end{equation}           
Here the upper and lower sign corresponds to the spectral function 
($\gamma$=$R$) and the occupied density of states ($\gamma$=$<$), 
respectively. The occupied density of states are closely related to the intensity 
in time-dependent photo-emission spectroscopy.\cite{PhysRevLett.102.136401} Note that in 
non-equilibrium, we can define the Fourier transform of a function $G(t,t')$ with 
respect to time-difference in forward ($t'>t)$, backward ($t'<t)$, or symmetric 
fashion. Since we analyze spectra at the largest time when the system is almost 
in a steady state, the differences are minor, and we have chosen the 
backward form \eqref{ftspec} unless stated otherwise.

In the left panels of Fig.~\ref{fig:fig4}(a) and (b), we compare the 
time-dependent spectral function in the photo-doped state (green solid lines) 
and in the equilibrium state (magenta dashed lines) for $\Delta=0.4$ 
[Fig.~\ref{fig:fig4}(a)], $\Delta=0.6$ [Fig.~\ref{fig:fig4}(b)], and different 
values of $U$. The corresponding occupation functions are shown in the right 
panels. In the thermal spectra, we find a  well-defined gap at small values 
of $U$. The gap is robust even though the temperature is larger than $U$ 
and $\Delta$, i.e., the main effect of the temperature is the occupation of 
states in the upper band, as seen from the occupied density of states in the 
right panel. Increasing values of $U$ lead to a renormalization of the gap, and 
a broadening of the square-root singularity at the gap edge.  When $U$ is of the 
order of the  ionic potential, the gap in the spectral function starts to fills, 
until it is completely melted for large $U$. Hubbard bands would only emerge 
at larger values of $U$. Similar as for the double occupancy, a significant difference 
between thermal and photo-excited systems in $A^<(\omega)$ is apparent only for 
$U\lesssim\Delta$, confirming the non-thermal nature of the photo-excited state in this 
regime. Interestingly, the difference between thermal and photo-excited states is 
almost invisible for the retarded spectral function on the energy scale plotted 
in the left panel of Fig.~\ref{fig:fig4}(a) and (b), which will be discussed 
below (Sec.~\ref{sec:discussion:a}).

\begin{figure}[t]
\centering
\includegraphics[angle=0,width=0.83\columnwidth]{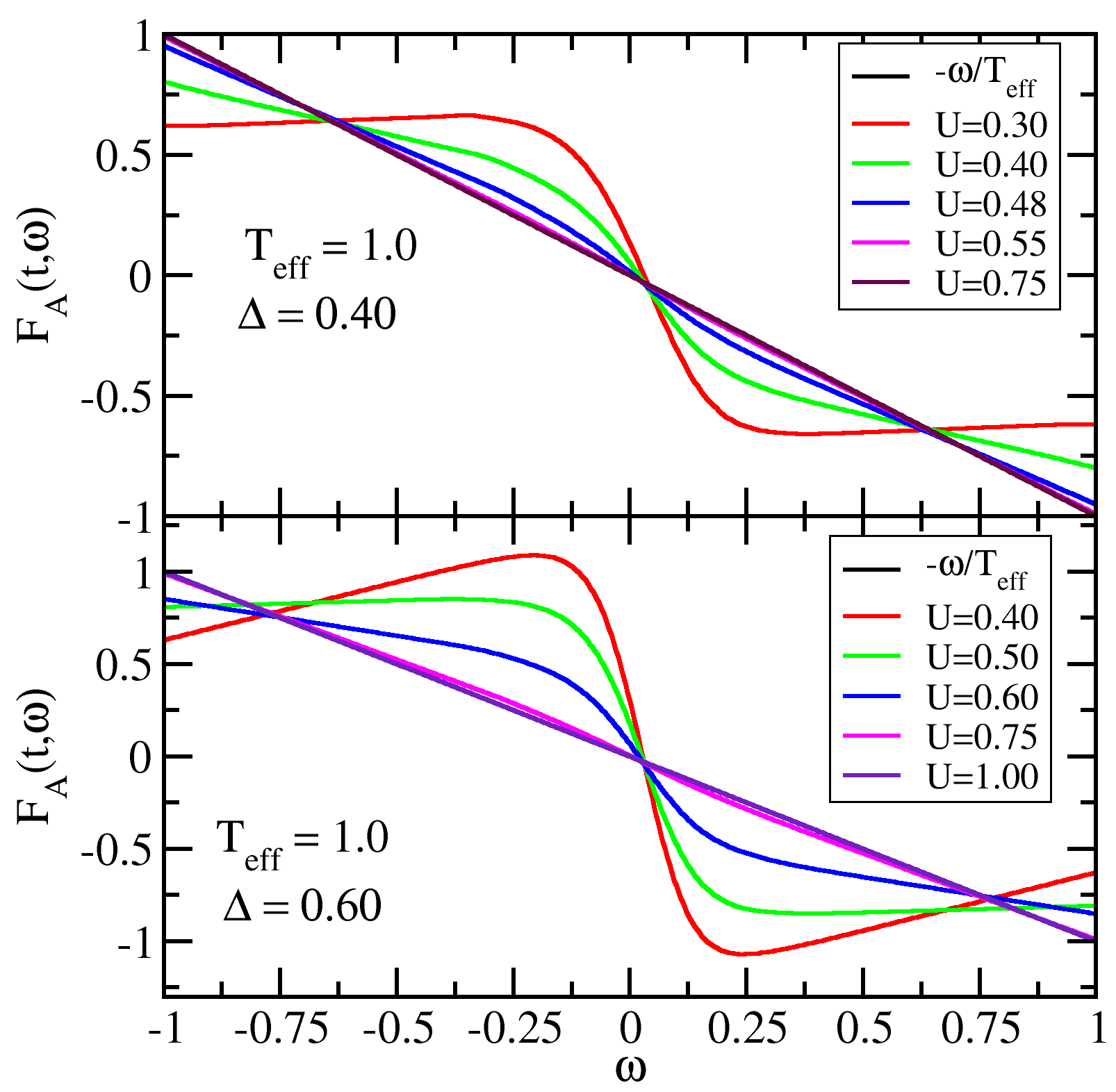}
\caption{Electronic distribution functions $F_{{A}}{({t=180},\omega)}$ 
[Eq.~\eqref{dist}] for different values of $U$ and ionic 
potential $\Delta$ =0.4 (upper panel) and $\Delta$=0.6 (lower panel).}
\label{fig:fig5}
\end{figure}

The thermalization can be further verified by checking whether the electronic 
Green's functions satisfy the fluctuation-dissipation theorem (FDT). In 
equilibrium, the FDT implies a ratio $A^<(\omega)/A^>(\omega)=e^{-\beta(\omega-\mu)}$ 
between the occupied density of states $A^<(\omega)=-iG^<(\omega)$ and the 
unoccupied density of states $A^>(\omega)=iG^>(\omega)$, or equivalently a 
ratio $A^<(\omega)/A^R(\omega)=1/(1+e^{\beta(\omega-\mu)})$ between the occupied 
density of states and the spectral function. To test whether the photo-doped state 
is in a quasi-equilibrium state, we therefore calculate the partial Fourier 
transform $A^{<,>}(t,\omega)$ as in Eq.~\eqref{ftspec}, and evaluate the quantity 
\begin{equation}
%\begin{align}
F_A(t,\omega)=\log\Big(\frac{A^<(t,\omega)}{A^>(t,\omega)}\Big),
\label{dist}
\end{equation} 
which reduces to $-(\omega-\mu)/T_{\text{eff}}$ in a quasi-equilibrium 
state. In Fig.~\ref{fig:fig5}, we plot $F_A(t,\omega)$ for a given 
$\Delta$ and different values of $U$. The FDT is satisfied when the local 
Coulomb interaction is greater than the ionic potential $U>\Delta$, which confirms 
that the photo-excited electrons have reached an equilibrium state. Furthermore, in 
this limit the effective temperature obtained from the FDT analysis accurately 
matches the value $T_\text{eff}=1$ obtained from the total energy. 

For $U\lesssim\Delta$, however, the electronic distribution functions are 
highly non-thermal. Nevertheless, $F_A(t,\omega)$ takes a linear 
form ($a+b\omega$) in the spectral region outside the gap 
($|\omega| \gtrsim 0.3$). This can be interpreted in the form of two distinct 
Fermi functions with different chemical potentials in the 
upper and lower band, respectively. This signals thermalization of 
electrons and holes as two separate sub-systems due to intra-band 
scattering. The effective temperatures of such photo-doped 
states are very large (the distribution functions are almost flat) or 
even negative, corresponding to a population inversion. In photo-doped Mott 
insulators with long-range Coulomb interactions, self-sustained population 
inversions have been found in the photo-doped state.\cite{PhysRevLett.118.246402} In 
the present case, however, the effective temperature is mainly governed by the 
carrier distribution during the photo-doping process.

\begin{figure}[t]
\centering
\includegraphics[angle=0,width=0.83\columnwidth]{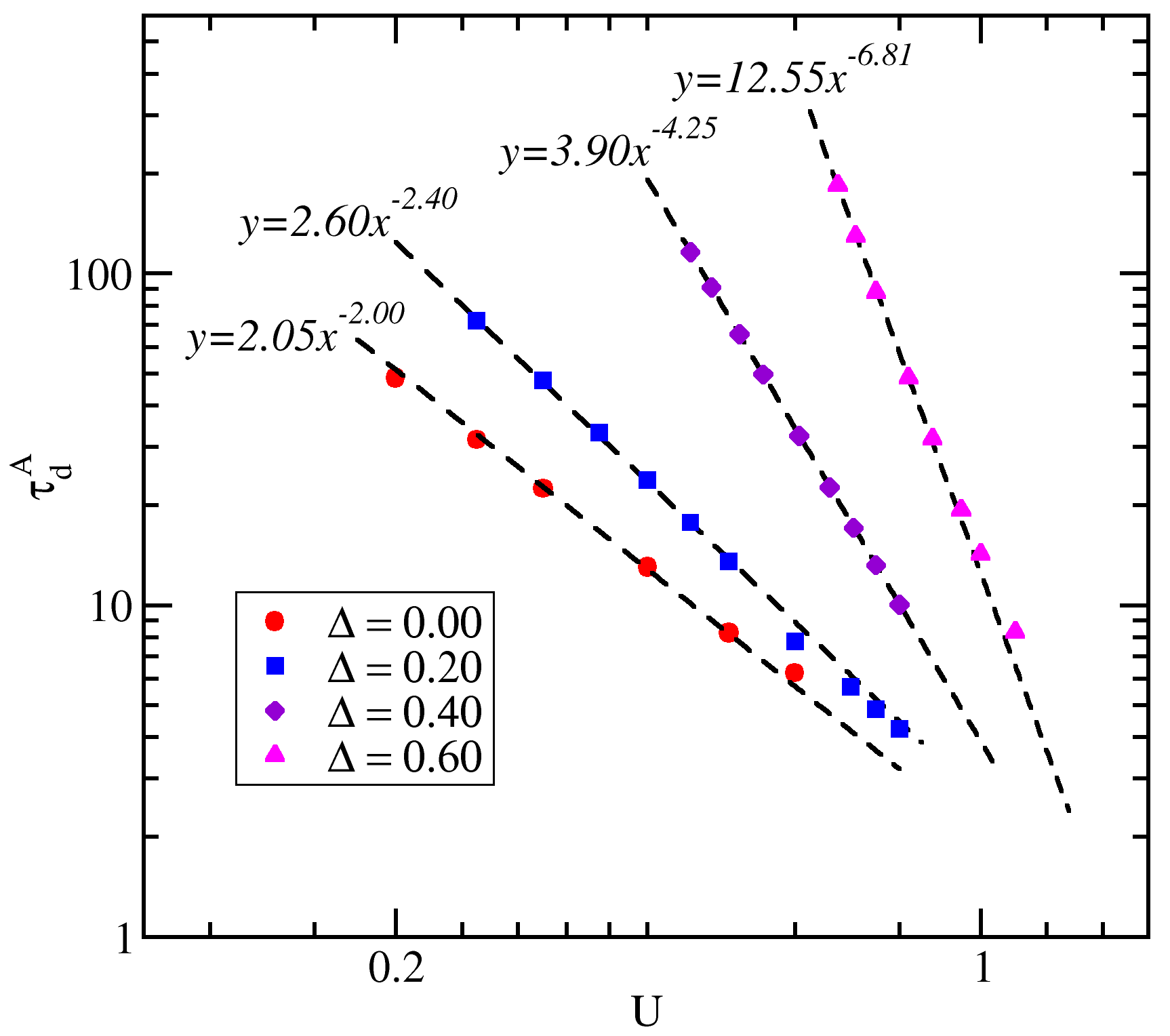}
\caption{Relaxation times of the double occupancy for different 
values of $\Delta$ and $U>\Delta$. The dashed curves indicate power law fits.}
\label{fig:fig6}
\end{figure}

\begin{figure}[t]
\centering
\includegraphics[angle=0,width=0.83\columnwidth]{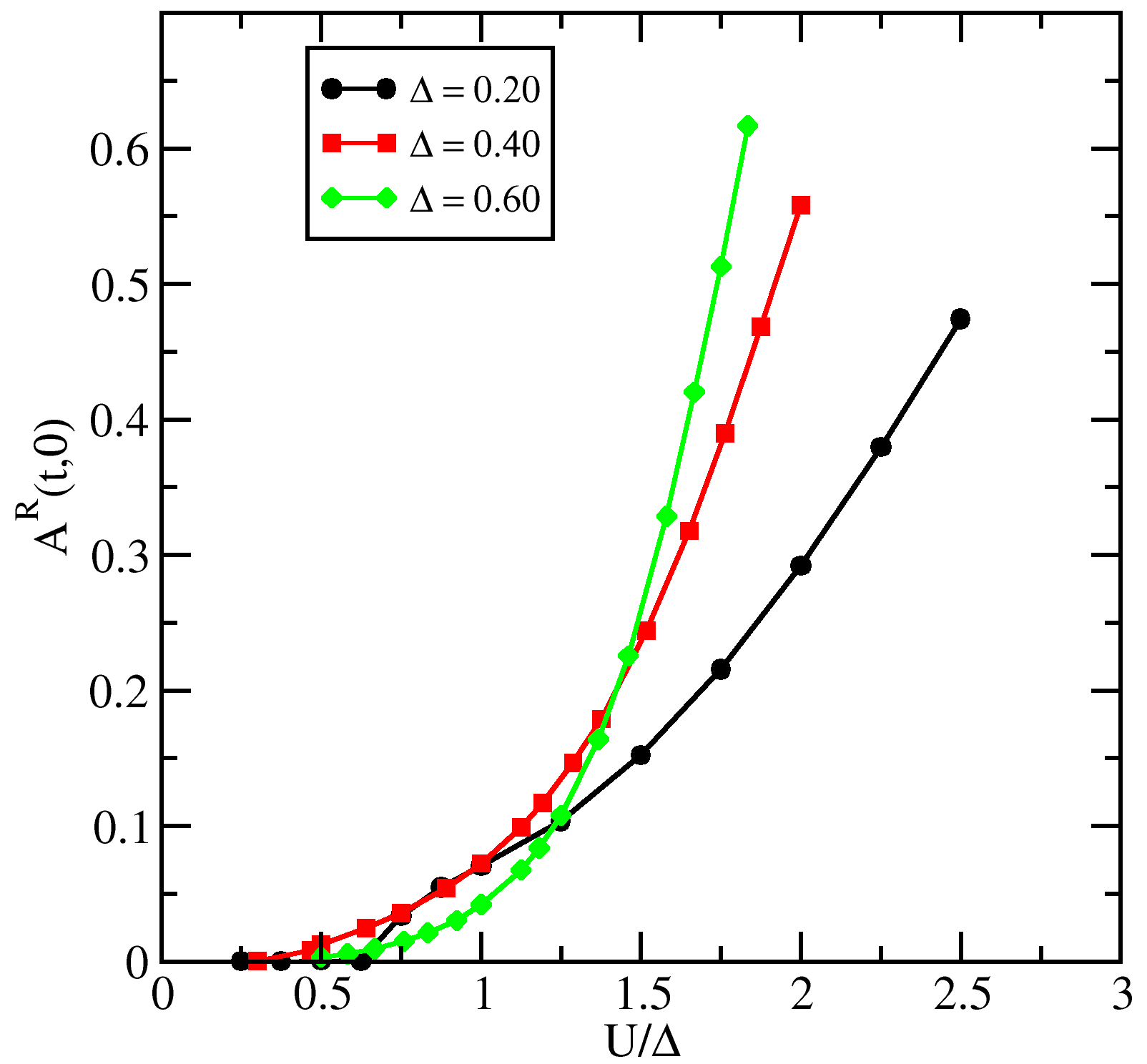}
\caption{The value of the retarded spectral function at $\omega$=0 for large 
time $t$=180 in the photo-excited state (excitation density such that $T_{\rm{eff}}$ = 1).}
\label{fig:fig7}
\end{figure}

\section{Discussion}
\label{sec:discussion}

\subsection{Crossover from a band-insulator to a correlated band insulator}
\label{sec:discussion:a}

Summarizing the previous section, one can say that the 
behavior of the system for $U\lesssim \Delta$ is well described by the 
expectation for a band insulator, which is characterized by the following features: 

(i) At temperatures of interest for photo-excitation, the (retarded) spectral 
function is relatively rigid. Up to a trivial Hartree shift, which depends only 
on the thermal or photo-induced sub-lattice occupation, it is barely influenced by increasing 
the temperature or photo-doping. This is seen by a comparison 
of thermal, photo-excited, and Hartree spectra in Fig.~\ref{fig:fig4}. The fact that 
the spectra only weakly depend on the occupation also explains why the difference 
between thermal and photo-excited spectra is small in this regime. 

(ii) The main relaxation mechanism is intra-band relaxation, i.e., scattering 
between electrons and holes. Only few scattering 
processes lead to a redistribution of carriers across the gap. This in 
consistent with the establishment of separate thermalized distributions in the 
region of the upper and lower band, as shown in Fig.~\ref{fig:fig5}. The absence 
of a global thermal state with a common chemical potential also explains the 
non-thermal behavior observed in the double occupancy and the sub-lattice 
occupation, which are quantities that involve both valence and conduction band 
states and thus can indicate global thermalization. 

%\subsection{Comparison to the metal}

\begin{figure*}[t]
\centerline{
\includegraphics[angle=0,width=0.85\textwidth]{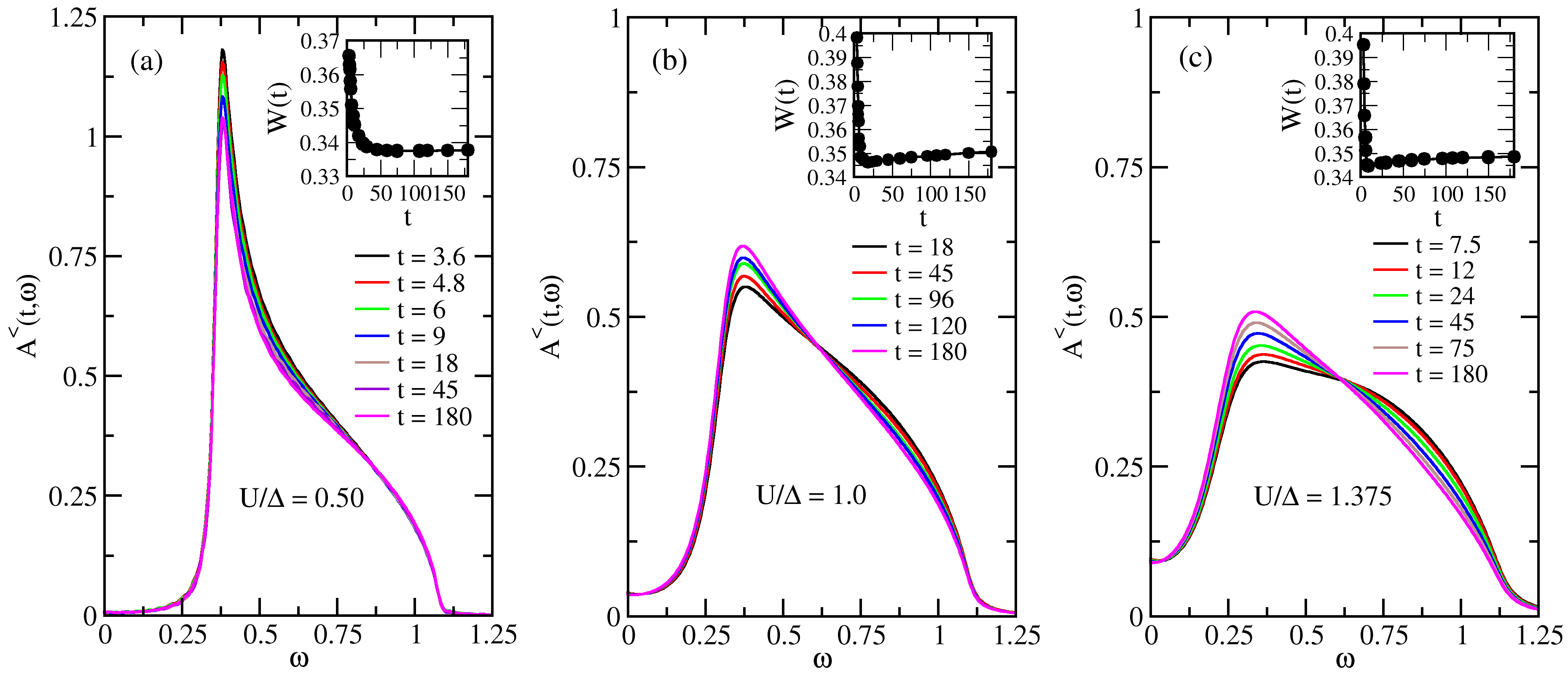}
}
\caption{Occupied density of states above the Fermi-level at 
different times for $\Delta$=0.4 and $T_{\rm{eff}}$=1.0. Insets: 
Integrated weight of the occupied density of states, $W(t)$=$\int^{\infty}_0 A^<(t,\omega) d\omega$
 above the Fermi-level.
}
\label{fig:fig8}
\end{figure*}

\begin{figure}[t]
\centering
\includegraphics[angle=0,width=0.83\columnwidth]{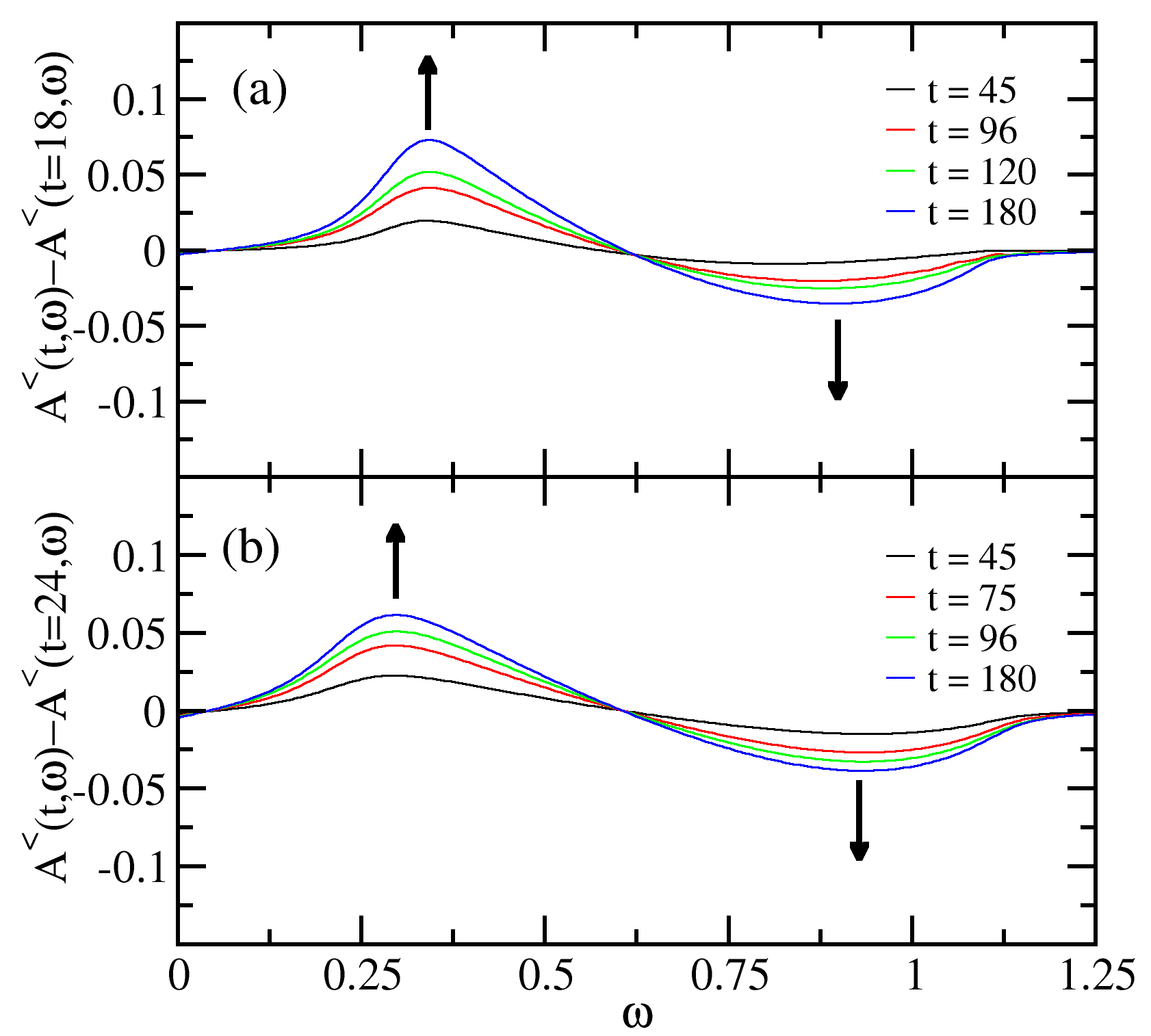}
\caption{Time dependent change of the occupied density of states for 
excitation density corresponding to $T_{\rm{eff}}$=1.0, $\Delta$=0.4, and (a) $U/\Delta$ = 
1.0 (b) $U/\Delta$ = 1.375. The up (down) arrows indicates frequencies at which 
rapid increase (decrease) most strongly during the relaxation.}
\label{fig:fig9}
\end{figure}

Above the crossover $U \approx \Delta$, the behavior of the system changes 
drastically. In this regime, photo-excitation leads to a partial filling-in or 
closing of the gap, and rapid thermalization. As the system thermalizes in the 
correlated band insulating regime, it is interesting to compare the thermalization 
times from Figs.~\ref{fig:fig2} and \ref{fig:fig3} with those observed in 
the metal (Fig.~\ref{fig:fig1}). In Fig.~\ref{fig:fig6}, we plot the 
thermalization times of the double occupancy 
for different values of the ionic potential and $U>\Delta$. We find that the decrease of 
the thermalization time with $U$ in the correlated band insulator is faster compared to the 
metal. The thermalization timescale is rather well described with a behavior $1/U^n$ with 
$n>2$ ($n=2$ for the metal), and the value of $n$ increases rapidly with the ionic potential. 
The large value of $n$ signifies that the thermalization dynamics of correlated band 
insulators is different from metals.

In Fig.~\ref{fig:fig7} we show the spectrum $A^R(0)$ at the Fermi-level as a 
function of $U/\Delta$ for different values of $\Delta$. A nonzero density of states 
at the Fermi-level gives a finite phase-space volume for photo-doped carriers to exchange 
the energy and thus can lead to fast thermalization of the photo-excited system. For 
$U\gtrsim \Delta$, the weight at the Fermi-level increases strongly with $U/\Delta$. The 
larger exponent $n$ in Fig.~\ref{fig:fig6} is therefore a consequence of the combined 
increase of the interaction $U$ and the spectral weight.

\subsection{Role of Impact ionization}

Recently it was shown that impact ionization processes can take an 
important role in the thermalization and relaxation of photo-excited Mott 
insulators.\cite{PhysRevB.90.235102} Impact ionization occurs if the kinetic energy 
of charge carriers is larger than the size of the Mott gap, so that it is 
energetically allowed to create an additional doublon-hole pair via 
two-particle scattering. Impact ionization can be identified by the following 
characteristic signatures in the occupation function: (i) An impact ionization 
process decreases the occupied density of states at high energies, and increases it 
at energies which are lower by at least the size of the gap. (ii) Impact 
ionization processes increase the density mobile carriers, and hence increase of 
integrated spectral weight above the gap. 

To see whether impact ionization processes are important in the thermalization of 
correlated band insulators, we plot the time-dependent lesser spectral function 
for three different ratios of $U/\Delta$=0.5, 1.0 and 1.375 (see Fig.~\ref{fig:fig8}). 
For $U/\Delta=0.5$ [Fig.~\ref{fig:fig8}(a)], the photo-excited system is not yet 
thermalized and there is a well-defined gap in the retarded spectral function. In this 
regime the weight of the occupied density of states decreases at the lower-edge of the 
conduction band, but the weight at upper-edge of the conduction band does not change at all, 
which is inconsistent with impact ionization. In the inset of Fig.~\ref{fig:fig8}(a), we 
show the total occupation above the Fermi-level as a function of time. 
After photo-excitation, the integrated spectral weight decreases rapidly with time and 
reaches saturation at short time scales. We do not observed an increase at later times, 
which shows that impact ionization processes are not relevant on 
the timescale of our simulation when the local Coulomb interaction is smaller than ionic potential.

In Fig.~\ref{fig:fig8}(b), we consider the case where the ionic potential is 
comparable to the Coulomb interaction ($U/\Delta=1.0$). In this case, there is already 
some spectral weight at the Fermi-level. One can observe that the occupied density of states increases at the 
lower-band edge of the conduction band and decreases at the upper-band edge decreases, and 
the overall integrated spectral weight above $\omega=0$ increases at later times 
(inset of Fig.~\ref{fig:fig8}). Both signatures indicate impact ionization processes 
take place in this regime. Similarly, we can also identify signatures of impact ionization 
process when $U$ is greater than the ionic potential as shown in Fig.~\ref{fig:fig8}(c). 

The precise time-dependent change of the occupied density of states can be used to 
quantify the significance of impact ionization process at later times in the 
relaxation process. In Fig.~\ref{fig:fig9}, we show the difference of $A^<(t,\omega)$ 
between large times and some initial time $t=18$ for $U/\Delta=1.0$ and $t=24$ for $U/\Delta=1.375$. 
The time-dependent change of the occupied density of states is positive for frequencies 
below $\omega \sim$ 0.625, and negative above 0.625. The ratio of the increase of the 
weight below $\omega$ = 0.625 to the decrease of weight above $\omega$ = 0.625 can 
therefore give an estimate of number of mobile carriers produced in impact ionization 
processes. We find that this ratio is roughly 1.4 for $U/\Delta=1.0$ and 1.25 for 
$U/\Delta=1.375$ whereas relaxation only via impact ionization would suggest a ratio 
of three.\cite{PhysRevB.90.235102} This implies that impact 
ionization processes are less significant in correlated band insulators than in small 
gap Mott insulators. Probably multi-particle scattering mechanisms are also 
significant in the rapid thermalization of correlated band insulators.

\section{Conclusion}

In conclusion, we have studied the photo-excitation dynamics of correlated band 
insulators in the ionic Hubbard model. Depending on the ratio of the 
interaction $U$ and the ionic potential $\Delta$, we observe a qualitatively 
different behavior: For $U\lesssim\Delta$, the spectrum itself is weakly 
influenced by the excitation (apart from a slight photo-induced screening 
of the gap, which is described by the Hartree shift of the bands.) The 
relaxation of the system in this regime is characterized by intra-band 
carrier scattering, leading to a non-thermal intermediate state with 
separate thermal distributions of electrons and holes. Above a crossover 
$U\approx\Delta$, the behavior changes. Photo-excitation can lead to a rapid 
renormalization of the spectrum, a filling-in of the gap, and fast thermalization. 
This rapid thermalization of a small gap insulator is reminiscent of the 
thermalization of small gap Mott insulators,\cite{PhysRevB.84.035122} but in 
contrast to Mott insulators impact ionization processes\cite{PhysRevB.90.235102} are 
less significant. The strong renormalization of the spectral function indicates that 
the dynamics of the correlated band insulator in this regime is no longer well described by 
mere quasi-particle scattering in rigid bands, which would be captured by kinetic equations.

In equilibrium, two systems with an identical gap in the spectral function 
can have very different ratios $U/\Delta$. For example, an insulator with 
$U/\Delta\gtrsim 1$ and a gap $\Delta_*$ can be compared to an ideal band 
insulator with ionic potential $\Delta_*$ and $U\ll\Delta_*$. The present analysis 
shows how these two systems with very similar equilibrium single particle properties 
can be distinguished by their dynamical behavior. This may be used to classify 
weakly interacting insulators as either band-insulators or correlated band insulators. 
In equilibrium, a qualitative distinction of correlated band insulators and band insulators 
may be based on a different behavior of  spin and charge gaps,\cite{0953-8984-15-34-319,
PhysRevB.75.245122,PhysRevB.80.155116} and it will be interesting to see whether these 
pictures can be linked to the dynamical behavior. We are leaving these questions future 
work. The observed crossover from a non-thermal state to a thermal state can potentially 
be found in SrRu$_{1-x}$Ti$_x$O$_3$ and some of the 3d transition metal oxides with 
crystal field splitting.\cite{PhysRevB.76.165128}

\acknowledgments
We acknowledge fruitful discussions with Yusuf Mohammed, and financial support 
from the ERC Starting grant No. 716648. The calculations have been done at the 
PHYSnet cluster of the university of Hamburg.

\bibliographystyle{apsrev4-1}
\bibliography{apssamp}
\end{document}